\documentclass[aps,twocolumn,prl,superscriptaddress,showpacs,preprintnumbers,amsmath,amssymb]{revtex4-1}

\usepackage[pdftex]{graphicx}
\usepackage{epstopdf}
\usepackage{graphicx}
\usepackage{dcolumn}
\usepackage{bm}
\usepackage{amssymb}
\usepackage{wasysym}
\usepackage{amsmath} 
\usepackage{amsthm}
\usepackage{sidecap}
\usepackage{hyperref}
\usepackage{natbib}

\begin{document}

\def\Ef{$E_{\rm F}$}
\def\Eb{$E_{\rm B}$}
\def\Efmath{E_{\rm F}}
\def\Ed{$E_{\rm D}$}
\def\Tc{$T_{\rm C}$}
\def\kpara{k$_\parallel$}
\def\kparamath{{\bf k}_\parallel}
\def\kperp{{k}$_\perp$}
\def\Gbar{$\overline{\Gamma}$}
\def\Kbar{$\overline{K}$}
\def\Mbar{$\overline{M}$}
\def\Ed{$E_{\rm D}$}
\def\invA{\AA$^{-1}$}
\def\Ef{$E_{\rm F}$}
\def\Tc{$T_{\rm C}$}
\def\kpara{{k}$_\parallel$}
\def\kperp{{k}$_\perp$}
\def\dirGX{$\overline{\rm \Gamma}-\overline{\rm X}$}
\def\dirGY{$\overline{\rm \Gamma}-\overline{\rm Y}$}
\def\dirGK{$\overline{\rm \Gamma}-\overline{\rm K}$}
\def\pntG{$\overline{\rm \Gamma}$}
\def\pntX{$\overline{\rm X}$}
\def\invA{\AA$^{-1}$}
\def\G{$\Gamma$} 
\def\Z{$Z$}

\title{Strong spin dependence of correlation effects in Ni due to Stoner excitations}

\author{J. S\'anchez-Barriga}
\email[Corresponding author. E-mail address: ] {jaime.sanchez-barriga@helmholtz-berlin.de.}
\affiliation{Helmholtz-Zentrum Berlin f\"ur Materialien und Energie, Albert-Einstein-Str.~15, 12489 Berlin, Germany}
\author{R. Ovsyannikov}
\affiliation{Helmholtz-Zentrum Berlin f\"ur Materialien und Energie, Albert-Einstein-Str.~15, 12489 Berlin, Germany}
\author{J. Fink}
\affiliation{Leibniz Institute for Solid State and Materials Research Dresden, Helmholtzstr. 20, D-01069 Dresden, Germany}
\affiliation{Max Planck Institute for Chemical Physics of Solids, N\"othnitzerstr. 40, D-01187 Dresden, Germany}
\affiliation{Institut f\"ur Festk\"orperphysik, Technische Universit\"at Dresden, D-01062 Dresden, Germany}

\begin{abstract}
Using high-resolution angle-resolved photoemission, we observe a strong spin-dependent renormalization and lifetime broadening of the quasiparticle excitations in the electronic band structure of Ni(111) in an energy window of $\sim$0.3 eV below the Fermi level. We derive a quantitative result for the spin-dependent lifetime broadening by comparing the scattering rates of majority and minority $d$ states, and further show that spin-dependent electron correlations are instead negligible for $sp$ states. From our analysis we experimentally determine the effective on-site Coulomb interaction $U$ caused by Stoner-like interband transitions between majority and minority $d$ states. The present results unambiguously demonstrate the remarkable impact of spin-dependent electron correlation effects originating from single-particle excitations in a prototypical 3$d$ transition metal, paving the way for further refinement of current many-body theoretical approaches.
\end{abstract}
 

\maketitle

Many-body interactions are of crucial importance in solids, and ultimately determine their electronic properties \cite{Senatore1994}. For more than half a century, quasiparticle excitations, which lead to a renormalization of the electronic band structure, have been investigated both theoretically and experimentally \cite{Kevan1992}. During the last decade, the important experimental progress in photoelectron and related spectroscopies has given access to quasiparticle properties in an unprecedented detail \cite{Hufner2007}. This development has led to a better understanding of the strong influence of electron correlation effects in condensed-matter systems such as 3$d$ transition metals and their alloys \cite{Herring1966}, high-$T{_c}$ superconductors \cite{Damascelli2003}, or heavy-fermion semiconductors \cite{Gunnarsson1983}, among many important examples.

In the case of ferromagnetic 3$d$ tansition metals, such as Fe, Co and Ni, it has been long understood that a proper description of their electronic properties cannot be achieved without taking into account exchange and electron correlation effects \cite{Plummer1979,Braun1996,Manghi1999}. Both are in fact crucial ingredients playing a key role in the appearance of ferromagnetism. In consequence, it turned out that the experimentally observed electronic structure of prototypical ferromagnets could not be properly described by calculations within the density functional theory (DFT) \cite{Hohenberg1964, Martin2004} in the local spin density approximation (LSDA) \cite{Jones1989}, which takes into account many-body interactions only partially. Over the last few years, a much better agreement between theory and experiment has been achieved for Fe, Co and Ni due to important developments in current theoretical approaches such as DFT plus dynamical mean field theory (DMFT) \cite{Braun2006,Kotliar2006,Rohringer2018} or the three-body scattering approximation (3BS) \cite{Greber1997,Monastra2002,Manghi2015}. Both schemes go beyond LSDA and provide a much more accurate description of many-body interactions, i.e, electron correlation effects, which are represented by a complex self-energy function $\Sigma(E,k)$ that can be directly accessed experimentally by means of angle-resolved photoemission (ARPES) \cite{Hufner2007,Damascelli2003}. Here the real part Re$\Sigma$ is related to the mass enhancement and the imaginary part Im$\Sigma$ to the scattering rate or the inverse quasiparticle lifetime.

One of the most important outcomes of 3BS and DMFT is a strong spin-dependent renormalization of the quasiparticle bands \cite{Monastra2002,Grechnev2007}. This effect is qualitatively predicted and detected at binding energies of few eV away from the Fermi level, and could explain, e.g., the experimentally observed quenching of the majority spin excitations in Co \cite{Monastra2002}. Conversely, a more detailed comparison between experiments and theory for Fe, Co and Ni revealed that state-of-the-art many-body calculations are not sufficient to reach quantitative agreement, especially concerning the mass renormalization and the spin dependence of the scattering rates \cite{Sanchez-Barriga2009,*Sanchez-Barriga2010,*Sanchez-Barriga2012}. This finding is remarkable in particular for Ni, where electron correlations are predicted to be stronger within the 3$d$ series, as evidenced by an on-site Coulomb interaction $U$ reaching theoretical values of $\sim$3 eV \cite{Katsnelson2002}. The reason is that large $U$ values enable 3BS and DMFT calculations to better reproduce the experimentally observed width of the occupied Ni 3$d$ bands, their reduced exchange splitting, as well as the Ni satellite appearing at $\sim$6 eV due to correlation effects \cite{Braun2006, Manghi1997}. 

Photoemission experiments on Ni \cite{Higashiguchi2005,Hofmann2009} have also revealed the importance of many-body interactions on a binding energy scale smaller than the Coulomb and exchange interactions. This conclusion, which is indeed at the focus of current theoretical approaches \cite{Grete2011}, mainly concerns the impact of electron-phonon \cite{Higashiguchi2005} and electron-magnon \cite{Hofmann2009} interactions in the electronic structure of Ni near the Fermi level. However, the central and most important question of up to which extent spin-dependent electron correlation effects in Ni are also important, including the observation of the principal mechanism underlying this phenomenon, have remained elusive so far.

Therefore, in this work we experimentally investigate the spin-dependent renormalization and scattering rate of the quasiparticle excitations in the electronic band structure of Ni near the Fermi level. Our main finding is that at binding energies below $\sim$0.3 eV the quasiparticle lifetimes of Ni $d$ states are strongly spin dependent due to the opening of new excitation channels related to single-particle excitations, while the effect is negligible for $sp$ states. The present results might serve as a unique benchmark for further refinement of many-body theoretical approaches that possibly include non-local fluctuations in single band systems.

\begin{figure}
\centering
\includegraphics [width=0.29\textwidth]{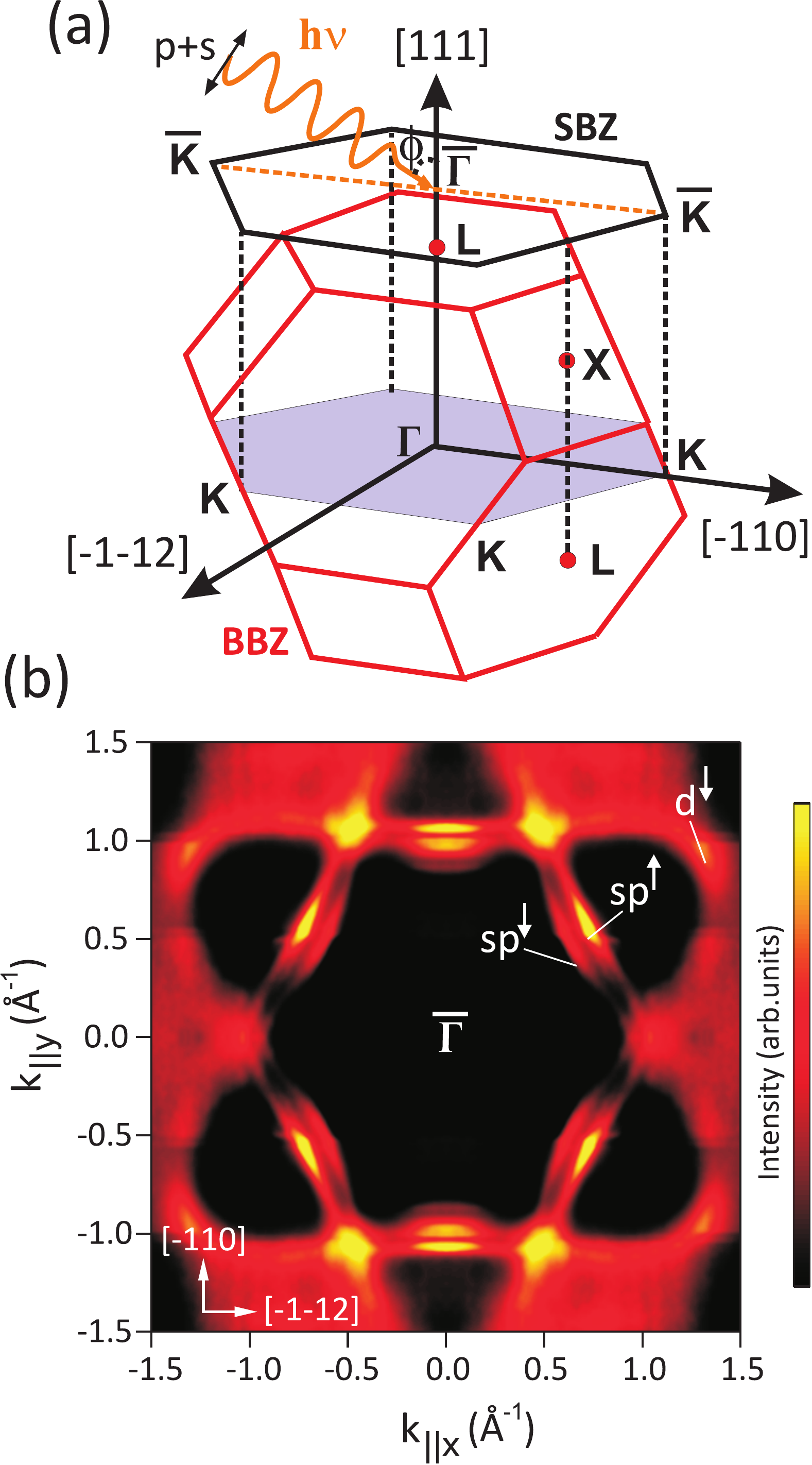}
\caption{(Color online) (a) Bulk and surface Brillouin zone of Ni(111). The light impinges the sample under an angle of $\phi=45^{\circ}$ with respect to the surface normal. The electron detection plane is oriented along the \Gbar\Kbar\ direction. (b) Fermi surface of Ni(111), containing clear contributions from majority ($\uparrow$) and minority ($\downarrow$) $sp$ and $d^{\downarrow}$ bands.}
\label{Fig1}
\end{figure} 
We performed high-resolution ARPES experiments at a temperature $T=$40 K using linearly-polarized undulator radiation at the UE112-PGM2 beamline of the synchrotron BESSY II. Photoelectrons were detected with a Scienta R8000 electron analyzer using the sample geometry shown in Fig.~\ref{Fig1}(a). The base pressure of the experimental setup was better than $1\cdot10^{-10}$ mbar. The Ni(111) surface was prepared on W(110) by deposition of 20 monolayers Ni and post-annealing. The structural quality of the film was verified by low-enery electron diffraction, and the sample was remanently magnetized along the [-110] direction. Overall experimental resolutions were set to 10 meV (energy) and $0.3\,^{\circ}$ (angular).

\begin{figure}
\centering
\includegraphics [width=0.35\textwidth]{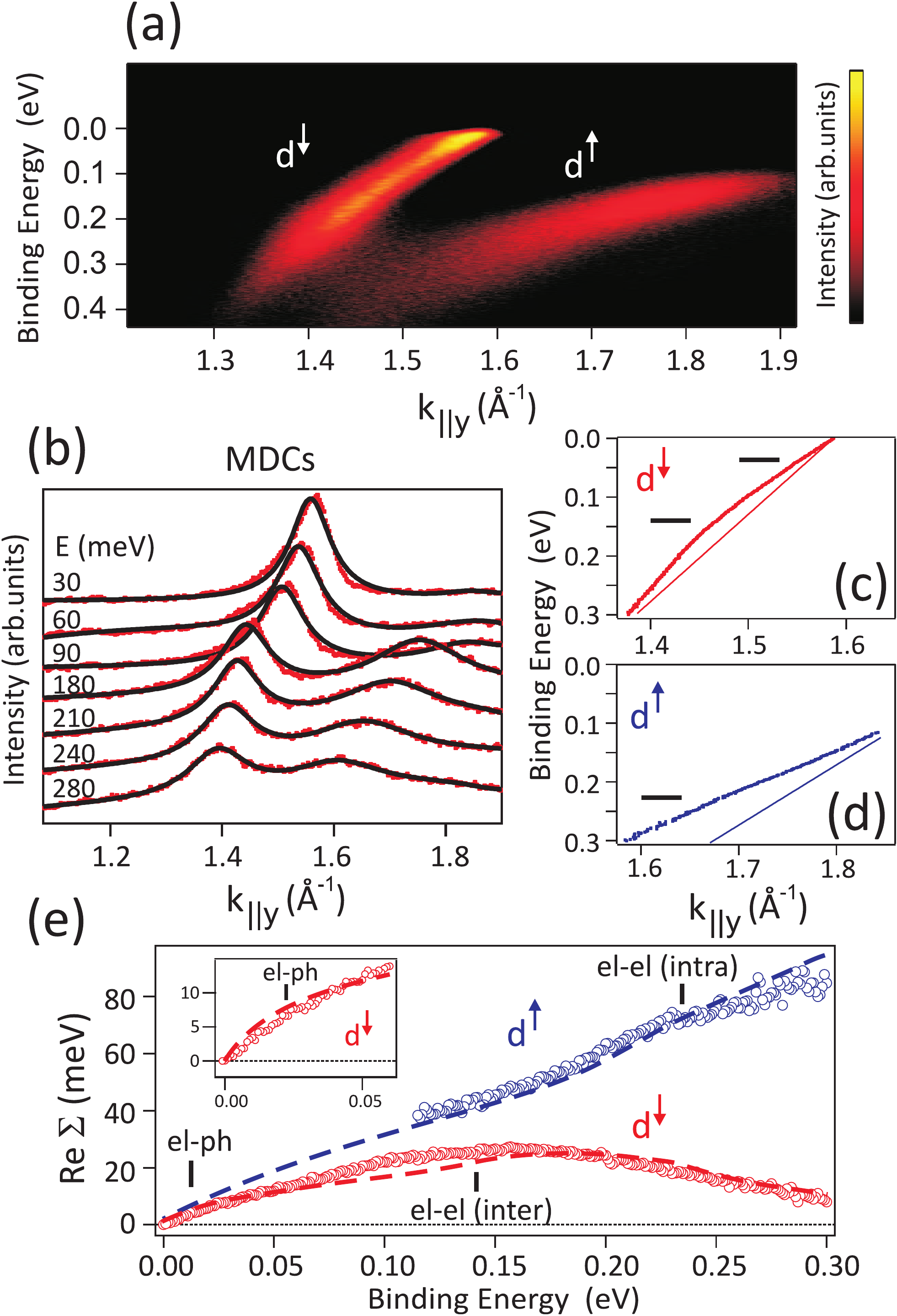}
\caption{(Color online) (a) High-resolution ARPES dispersions of the $d$ bands of Ni(111). (b) Selected fits [black (dark) solid lines] to experimental MDCs [red (light) dotted lines] as a function of binding energy. (c), (d) MDC-derived band dispersions for (c) $d^{\downarrow}$ and (d) $d^{\uparrow}$ bands. The bare particle bands are shown as solid lines. The kinks are highlighted by horizontal black (dark) solid lines. (e) Re$\Sigma$ for $d^{\downarrow}$ [red (light)] and $d^{\uparrow}$ [blue (dark)] bands.  The kink structures are properly described (dashed lines) based on electron-phonon and electron-electron interactions. Inset: Zoom-in on the region near the Fermi level. }
\label{Fig2}
\end{figure} 
Figure~\ref{Fig1}(b) displays the Ni Fermi surface measured with 136 eV photons. At this photon energy, we cut the $k_{z}$ plane corresponding to the $\Gamma$ point of the bulk Brillouin zone (BBZ) \cite{Plummer1979}, as highlighted by a violet (dark) colored plane in Fig.~\ref{Fig1}(a). Therefore, under this condition, the $K$ point of the BBZ directly projects on the \Kbar\ point of the surface Brillouin zone (SBZ). The Fermi surface in Fig.~\ref{Fig1}(b) contains clear contributions from majority $sp^{\uparrow}$ and minority $sp^{\downarrow}$ states, as well as from the minority $d^{\downarrow}$ band which is partially unoccupied \cite{Greber1997}. The spin assignment of the different bands is fully consistent with previous spin-resolved measurements \cite{Hoesch2002}. Because the ferromagnetic splitting of $sp$ states is rather small, the shape of the two exchange-split $sp$ sheets resembles the Fermi surface of Cu \cite{Aebi1994}, despite their significant hybridization with $d^{\downarrow}$ states especially at the intersection regions near the projected bulk $L$ and $X$ points. On the other hand, the majority $d^{\uparrow}$ states are completely filled and thus there are no $d^{\uparrow}$ sheets contributing to the measured Fermi surface. 

\begin{figure}
\centering
\includegraphics [width=0.35\textwidth]{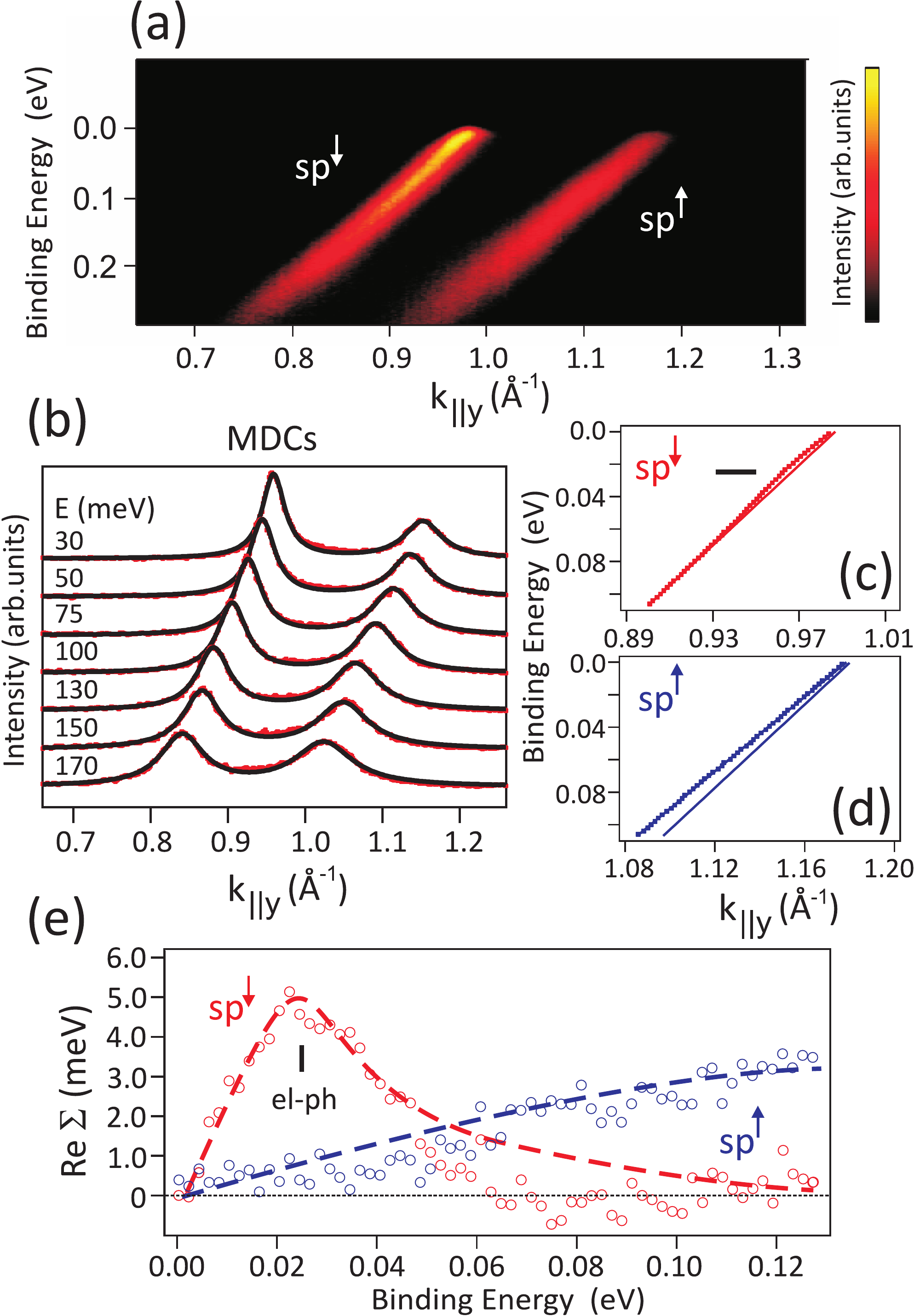}
\caption{(Color online) (a) High-resolution ARPES dispersions of the $sp$ bands of Ni(111). (b) Selected fits [black (dark) solid lines] to experimental MDCs [red (light) dotted lines] as a function of binding energy. (c),(d) MDC-derived band dispersions for (c) $sp^{\downarrow}$ and (d) $sp^{\uparrow}$ states. The bare particle bands are shown as solid lines. The kink in the $sp^{\downarrow}$ band is marked by an horizontal black (dark) solid line. (e) Re$\Sigma$ for $sp^{\downarrow}$ [red (light)] and $sp^{\uparrow}$ [blue (dark)] bands. It can be properly described (dashed lines) based on electron-phonon and electron-electron interactions.}
\label{Fig3}
\end{figure}

\begin{figure}
\centering
\includegraphics [width=0.4\textwidth]{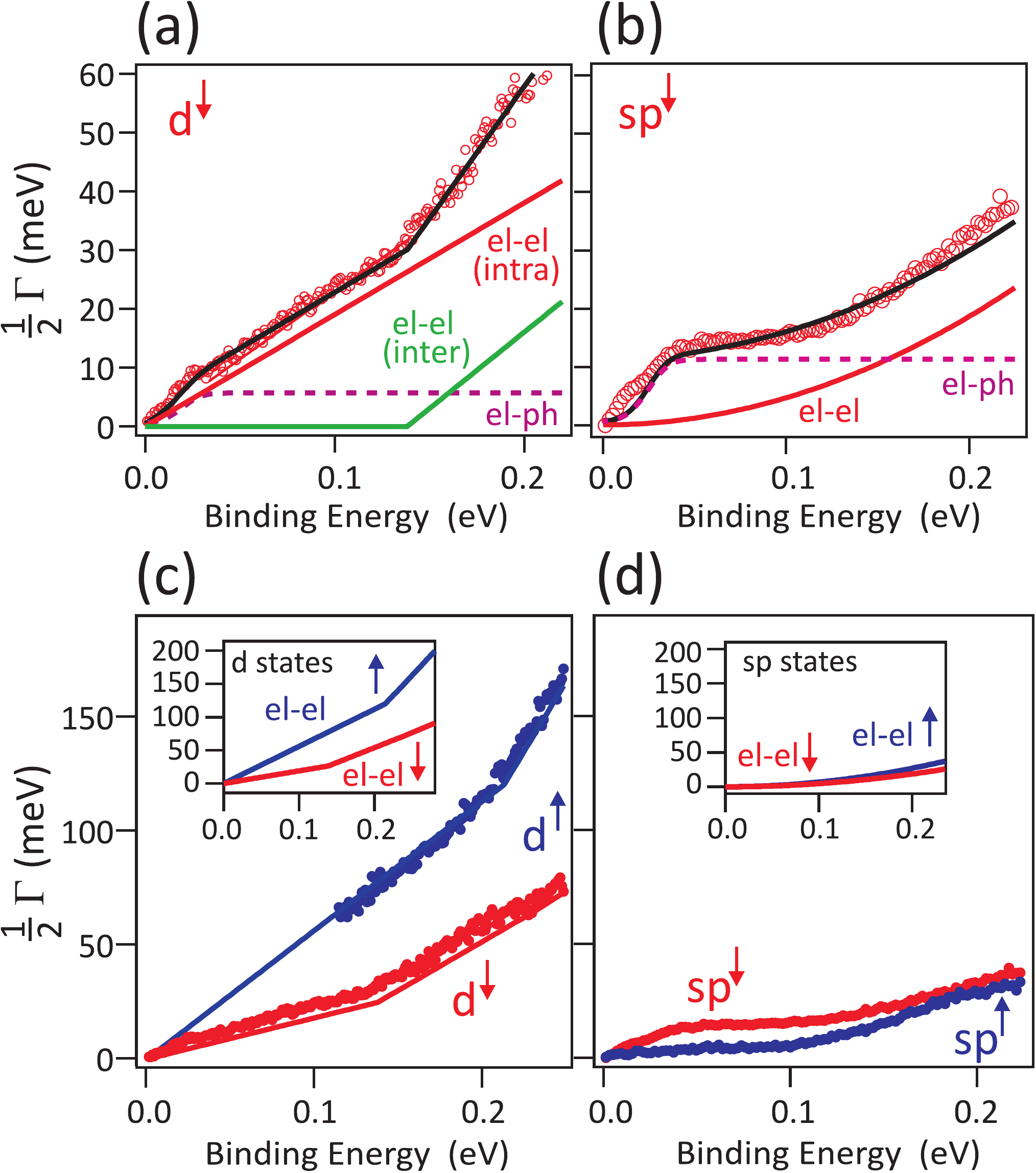}
\caption{(Color online) Scattering rates, represented as $\frac{1}{2}\Gamma$, of (a) $d^{\downarrow}$ and (b) $sp^{\downarrow}$ states. The contributions from electron-phonon and electron-electron scattering are shown separately, and their sum as black (dark) solid lines. (c),(d) Spin dependence of the scattering rates for (c) $d$ and (d) $sp$ states. Insets: Spin-dependent electron correlations for majority [blue (dark)] and minority [red (light)] states.}
\label{Fig4}
\end{figure}

To investigate the impact of electron correlation effects quantitatively, we perform a detailed analysis of the energy positions and spectral width of majority and minority $d$ and $sp$ bands, as shown in Figs.~\ref{Fig2} and \ref{Fig3}, respectively. In the dispersion and in the difference between the bare particle dispersion and the experimental dispersion one can clearly see kinks due to electron-phonon and electron-electron interaction which are marked by sticks and that will be discussed below. Measurements were taken along the \Gbar\Kbar\ direction of the SBZ, meaning that the energy-momentum dispersions shown in Figs.~\ref{Fig2}(a) and \ref{Fig3}(a) cut perpendicularly the constant-energy contours of the different states up to the Fermi level in Fig.~\ref{Fig1}(b). In this way, we are able to accurately fit the experimental momentum-distribution curves (MDCs) with Lorentzians on a linear background after convolution with a Gaussian function representing the momentum resolution, as shown in Figs~\ref{Fig2}(b) and \ref{Fig3}(b). The corresponding MDCs-derived quasiparticle dispersions of $d$ and $sp$ states are shown in Figs.~\ref{Fig2}(c)-\ref{Fig2}(d) and \ref{Fig3}(c)-\ref{Fig3}(d), respectively. We firstly observe several kink structures appearing in the minority $d$ and $sp$ bands when comparing each MDC-derived dispersion to the corresponding LDA bare particle bands \cite{Goldmann1999,Higashiguchi2005,Hofmann2009,Buenemann2003}. In Figs.~\ref{Fig2}(e)-\ref{Fig3}(e) we provide an estimate of Re$\Sigma$. It is clearly seen that for the $d$ bands the mass renormalization (i.e, Re$\Sigma$) is close to a factor of 2 larger for majority spin electrons. The corresponding scattering rates $\Gamma$ of the $d^{\downarrow}$ and $sp^{\downarrow}$ bands, represented as $\frac{1}{2}\Gamma$, are shown in Figs.~\ref{Fig4}(a) and \ref{Fig4}(b), respectively.

Note that $\frac{1}{2}\Gamma$ is directly related to the imaginary part of the complex self-energy as $\frac{1}{2}\Gamma\!=\!v^*\frac{1}{2}{\omega_k}\!=\!\frac{v^*}{v_b}$Im$\Sigma$, where $v^*$($v_b$) is the renormalized (bare) group velocity and $\frac{1}{2}{\omega_k}$ the corresponding half width at half maximum of the Lorentzian peaks. To understand the relationship between the experimental results shown in Figs.~\ref{Fig4}(a) and Fig.~\ref{Fig2}(e), as well as in Figs.~\ref{Fig4}(b) and \ref{Fig3}(e), we describe the scattering rate as $\Gamma\!=\!\Gamma_{el-ph}\!+\!\Gamma_{el-el}$, where each term represents the contribution from electron-phonon and electron-electron interactions, respectively. The results of the model are shown as dashed lines in Figs.~\ref{Fig2}(e) and \ref{Fig3}(e) and as solid lines in Fig.~\ref{Fig4}. Other contributions such as impurity scattering and final-state broadening add up in a constant and energy-independent offset that amounts to $\sim$15 meV and it is subtracted from the experimental data shown in Fig.~\ref{Fig4}. 

Within the model, the real and imaginary parts of the complex self-energy are related by the Kramers-Kronig transformation \cite{Hufner2007}. We describe the electron-phonon interaction in terms of the Eliashberg function for bosonic-like excitations as in the Debye model \cite{Hufner2007}. For the $d$ bands, the linear dependence as a function of binding energy in particular sections observed in Fig.~\ref{Fig4} is consistent with a non-Fermi liquid behavior that can be described by $\Gamma_{el-el}\sim\gamma E$ as for other strongly correlated systems \cite{Fink2015}. The different linear sections correspond to intraband and interband Stoner-like excitations. For the $sp$ bands, the contribution from electron-electron scattering is accounted for by a coupling strength $\beta$ and follows a quadratic dependence with binding energy according to the Fermi liquid theory \cite{Landau1956}. 

From this analysis, for the $d^{\downarrow}$ band we derive a mode energy of $E_{el-ph}$=26$\pm$5 meV associated to electron-phonon interaction with a coupling constant $\lambda_{el-ph}$= 0.37$\pm$0.03, and an energy of the higher-energy kink $E_{el-spin}$=136$\pm$20 meV associated to Stoner-like excitations. For the $sp^{\downarrow}$ band we derive $\lambda_{el-ph}$= 0.28$\pm$0.04 and a mode energy of $E_{el-ph}$=31$\pm$5 meV. These results agree only qualitatively with previous findings \cite{Higashiguchi2005,Hofmann2009}, where the fact that kinks were observed only in the minority bands remained unexplained. The difference in the values might be attributed to the different point in $k$ space measured here, indicating that non-local correlation effects \cite{Rohringer2018} in Ni cannot be completely neglected. One might argue that the magnitude of $\lambda$ for $sp$ states could be related to their relative $d$ weight \cite{Higashiguchi2005}, which is about 10\% smaller for the $sp^{\uparrow}$ band. However, this difference seems too small to explain the absence of an electron-phonon kink in the $sp^{\uparrow}$ band. The effect appears instead compatible with the spin dependence of electron-phonon coupling predicted for ferromagnetic metals \cite{Verstraete2013}. On the other hand, as the $d^{\uparrow}$ band is shifted up in energy due to the exchange splitting, $d^{\uparrow}$ states cannot host an electron-phonon kink because the exchange gap $\Delta$ is larger than $E_{el-ph}$. Furthermore, alternatively to an explanation in terms of coupling to magnons \cite{Hofmann2009}, the fact that $E_{el-spin}$ is close to the energy of the weakly-dispersing top of the majority $d^{\uparrow}$ band indicates that the high-energy kink in the $d^{\downarrow}$ band originates from an interband electron-hole excitation between the $d^{\uparrow}$ and $d^{\downarrow}$ bands together with a spin flip. A Stoner-like excitation is consistent with this picture, as also derived from neutron scattering results which indicate Stoner excitations near 120 meV and that collective spin excitations above that energy such as magnons are heavily Landau damped due to single-particle excitations \cite{MookPaul1985}. Such an interpretation is further supported by the fact that the majority $d^{\uparrow}$ band in Fig.~\ref{Fig2}(e) exhibits a kink at nearly twice higher energy, indicating that intraband transitions are responsible for this kink. The reason is that intraband electron-hole excitations within the majority $d^{\uparrow}$ band can only start at energies of the order of 2$\Delta$, because for smaller energies the relaxation of a hole by an energy $\Delta$ would require the excitation of an electron inside the gap which is not allowed \cite{Norman1998,Campuzano2001}. The fact that the $sp^{\uparrow}$ band crosses the Fermi level is also consistent with the lack of a high-energy kink as channel for Stoner-like excitations in the $sp^{\downarrow}$ band.

To further explore the existence of spin-dependent electron correlation effects, which up to date has remained elusive in particular for $d$ states, in Figs.~\ref{Fig4}(c) and \ref{Fig4}(d) we compare the scattering rates of majority and minority spin bands. This comparison is the central result of the present work. The scattering rate of the majority $d^{\uparrow}$ band in Fig.~\ref{Fig4}(c) changes slope at $E$=230$\pm$15 meV in agreement with the result of Fig.~\ref{Fig2}(e). We clearly observe that the lifetime broadening of $d$ states in Fig.~\ref{Fig4}(c) is strongly spin dependent, while the effect in Fig.~\ref{Fig4}(d) appears negligible for $sp$ states. This result is in qualitative agreement with the behavior expected from DMFT and 3BS theories \cite{Manghi2015,DiMarco2009}, according to which $sp$ electrons are less affected by (spin-dependent) electron correlations because they are much more delocalized and less influenced by the motion of ions. 

The insets of Figs.~\ref{Fig4}(c) and \ref{Fig4}(d) display the overall impact of spin-dependent electron correlations in the $d$ and $sp$ bands after discriminating the contribution from $\Gamma_{e-ph}$ to the measured scattering rates. From our analysis of the $sp$ bands we derive a spin-dependent electron-electron interaction strength of $\beta_{\uparrow}$=0.56$\pm$0.03 eV$^{-1}$ and $\beta_{\downarrow}$=0.48$\pm$0.03 eV$^{-1}$, while for the $d$ bands $\gamma_{\uparrow}$= 0.62$\pm$0.02 eV$^{-1}$ and $\gamma_{\downarrow}$=0.25$\pm$0.04 eV$^{-1}$. Interestingly, the averaged $\gamma$ value is comparable to those detected in nonmagnetic electron-doped ferropnictides, which are also supposed to be related to interband transitions \cite{Fink2015}. The strength of scattering processes of majority spin electrons, which involve the creation of minority spin pairs, is predicted to be proportional to $U$ \cite{Monastra2002,Manghi2015,Grechnev2007}. Similarly, scattering processes of minority spin electrons also involve the creation of minority spin pairs, but the effective interaction for parallel-spin electrons is predicted to be proportional to $U\!-\!J\!<\!U$ \cite{Monastra2002,Manghi2015,Grechnev2007}. Accordingly, from Fermi's golden rule \cite{Hufner2007} we derive $\gamma_{\uparrow}/\gamma_{\downarrow}=U^{2}/(U-J)^{2}$. Using the widely accepted atomic value for the averaged on-site exchange interaction $J$=0.9 eV \cite{Anisimov1991}, we obtain an experimental effective Coulomb interaction of $U_\text{eff}$=2.5$\pm$0.3 eV originating from interband transitions between $d^{\uparrow}$ and $d^{\downarrow}$ bands. This value agrees qualitatively with the one derived from theoretical studies \cite{Manghi2015,Grechnev2007,DiMarco2009}. The present results allow us to accurately determine the strength of spin-dependent electron correlations experimentally. While the effect is small for the $sp$ bands ($\beta_{\uparrow}\approx\beta_{\downarrow}$), for the $d$ bands $\gamma_{\uparrow}$/$\gamma_{\downarrow}$=2.5$\pm$0.2 is clearly visible without boson corrections. This is consistent with $\lambda^{\uparrow}_{el-el}/\lambda^{\downarrow}_{el-el}$=2.6$\pm$0.1 as derived from Re$\Sigma$. Our findings taken altogether unambiguously demonstrate the critical importance of spin-dependent electron correlation effects in the vicinity of the Fermi level of a typical 3$d$ transition metal, a conclusion which is also highly relevant for other correlated systems such as cuprates and ferropnictides \cite{Valla1999}. 

To summarize, by means of high-resolution ARPES we have investigated spin-dependent many-body effects in the electronic band structure of Ni near the Fermi level. We have analyzed various kink structures and provided a quantitative estimate of the strength of the coupling to bosonic-like and single-particle excitations affecting the majority and minority spin bands. Furthermore, we have derived a quantitative result for the spin-dependent lifetime broadening of the $d$ and $sp$ bands. Our results clearly demonstrate that the scattering rates of Ni $d$ states are strongly spin dependent, while the effect is less significant for $sp$ states. These findings unravel an unprecedented role of spin-dependent electron correlations originating from single-particle excitations in a prototypical 3$d$ transition metal. The present results demand for more refined many-body theoretical approaches that not only take into account on-site and off-site correlations, but that also fully capture the observed kink structures and strength of the spin-dependent electron correlations at a quantitative level.

\section {Acknowledgements}

J. F. thanks for the hospitality during his stay at BESSY II, Helmholtz-Zentrum Berlin.

\bibliographystyle{apsrev4-1}

\bibliography{Ni_biblio}

\end{document}